\documentclass[letterpaper,twocolumn,10pt]{article}
\usepackage{color,xcolor}
\usepackage{colortbl}
\usepackage{multirow}
\usepackage{amsmath}
\usepackage{booktabs}
\usepackage{url}
\usepackage{graphicx}
\usepackage{listings}
\usepackage{color}
\usepackage{multicol}
\usepackage{latexsym}

\usepackage{amsthm}\usepackage{lmodern}

\usepackage{mathpazo}

\usepackage{amsmath,latexsym,amssymb,amsthm,amsfonts}
\usepackage{xspace}%
\PassOptionsToPackage{hyphens}{url}\usepackage{hyperref}

\usepackage{algorithmicx}
\usepackage{algpseudocode}
\usepackage{algorithm}
\usepackage{macros}
\usepackage{ioa_code}

\title{{ Formal Verification of Blockchain Byzantine Fault Tolerance}}
\date{}

\author{Pierre Tholoniat\\{\small University of Sydney, Australia}\\{\small \'{E}cole Polytechnique, France} \and Vincent Gramoli\\{\small University of Sydney, Australia}\\{\small CSIRO Data61, Australia}}%

\begin{document}
\maketitle

\begin{abstract}
To implement a blockchain, the trend is now to integrate a non-trivial Byzantine fault tolerant consensus 
algorithm instead of the seminal idea of waiting to receive blocks to decide upon the longest branch.
After a decade of existence, blockchains trade now large amounts of valuable 
assets and a simple disagreement could lead to disastrous losses. Unfortunately, Byzantine consensus solutions used in blockchains are at best proved 
correct ``by hand'' as we are not aware of any of them having been formally verified.

In this paper, we propose two contributions:
\textit{(i)}~we illustrate the severity of the problem by listing six vulnerabilities of blockchain consensus including two new counter-examples;
\textit{(ii)}~we then formally verify two Byzantine fault tolerant components of Red Belly Blockchain using the ByMC model checker.
First, we specify a simple broadcast primitive in 116 lines of code that is verified in 40 seconds on a 2-core Intel machine. 
Then, we specify a blockchain consensus algorithm in 276 lines of code that is verified in 17 minutes on a 64-core AMD machine using MPI.

To conclude, we argue that it has now become both relatively simple and crucial to formally verify
the correctness of blockchain consensus protocols. %
\end{abstract}

\section{Introduction}

As blockchain is a popular abstraction to handle valuable assets, it has become one of the cornerstone of promising solutions for building critical applications without requiring trust.  
Unfortunately, after a decade of research in the space,  blockchain still appears in its infancy, unable to offer the guarantees that are needed by the industry to automate critical applications in production.
The crux of the problem is the difficulty of having remote computers agree on a unique block at a given index of the chain when some of them are malicious.
The first blockchains~\cite{Nak08} allow disagreements on the block at an index of the chain but try to recover from them before assets get stolen through double spending:
With disagreement, an asset owner could be fooled when she observes that she received the asset. Instead the existence of a conflicting block within a different branch of the chain may indicate that the asset belongs to a different user who can re-spend it.
This is probably why most blockchains now build upon some form of Byzantine consensus solutions~\cite{CNG18,BKM18,BG19}.

Solving the Byzantine consensus problem, defined 39 years ago~\cite{PSL80}, is needed to guarantee that machines agree on a common block at each index of the chain. The consensus was recently shown to be necessary in the general scenario where conflicting transactions might be requested from distributed servers~\cite{GKM19}. Various solutions to the consensus problem were proposed in the last four decades~\cite{CL02,KADCW09,SYB14,Kwo14,AGZ15,CGLR18,Lin19}.
Most of these algorithms were proved correct ``by hand'', often listing a series of Lemmas and Theorems in prose leading the reader to the conclusion that the algorithm solves agreement, validity and termination in all possible distributed executions. 
In the worst case, these algorithms are simply described with text on blog post~\cite{BBK18,Lin19}.
In the best case, a mathematical specification is offered, like in TLA+, but without machine-checked proofs~\cite{TS15}. 
Unfortunately, such a formal specification that is not machine-checked remains error prone~\cite{Sut19}.

As far we know, no Byzantine fault tolerant consensus algorithms used in blockchains 
have ever been formally verified automatically with the help of a program that produces an output to ascertain the specification correctness.
We do not claim that formally verifying blockchain consensus guarantees correct executions, as they are other tools necessary to execute it that may be incorrect. We believe instead that verifying blockchain consensus greatly reduces errors by 
forcing the distributed algorithm designer %
to write an automaton sufficiently disambiguated to be systematically evaluated with tools designed by verification experts.
While some consensus algorithms have been automatically proved correct~\cite{LWB17,BKL19}, 
these algorithms are mainly state-of-the-art algorithms. They do not necessarily offer the practical properties suitable for blockchains as they are not to be implemented in blockchains.

In this paper, we first survey important problems that recently affected blockchain consensus. 
In particular, we propose two new counter examples explaining why the Casper FFG algorithm, which  should be integrated in phase 0 of Ethereum 2.0 and the HoneyBadger, which is being integrated into one of the most popular blockchain software, called \texttt{parity},
may not terminate.
We also list four additional counter examples from the literature to illustrate the amplitude of the problem for blockchains.
While there exist alternative solutions to some of these problems that could be implemented it does not prevent other problems from existing. 
Moreover, proving ``by hand'' that the fixes solve the bugs may be found unconvincing, knowing that these bugs 
went unnoticed when the algorithms were proven correct, also ``by hand'', in the first place. 

We then build upon modern tools and equipments at our disposal to formally verify
blockchain consensus components that do not assume synchrony under the assumption that $t<n/3$ processes are Byzantine (or \emph{faulty}) among $n$ processes.
In particular, we explain how the Byzantine model checker ByMC~\cite{KW18} can be used by distributed computing scientists to verify blockchain consensus components without a deep expertise in formal verification.
The idea is to convert the distributed algorithm into a threshold automaton~\cite{LWB17} that
represents a state as a group of all the states in which a \emph{correct} (or non-faulty) process resides until 
this process receives sufficiently many messages to transition. 
We offer the threshold automaton specification of a Byzantine fault tolerant broadcast primitive that is key to 
few blockchains~\cite{MXC16,CGLR18,CGG19}.
Finally, we also offer the threshold automaton specification %
of a variant of the Byzantine consensus of Red Belly Blockchain~\cite{CGLR18} that we prove safe and live under the round-rigidity assumption~\cite{BKL19} that helps modeling a fair scheduler~\cite{BT85}, hence allowing other distributed computing scientists to reproduce the verification with this publicly available model checker.

Various specification languages (e.g.,~\cite{YPL99,Lyn03}) were proposed for distributed algorithms  before threshold automata, but they did not allow the simplification needed to model check algorithms as complex as the Byzantine consensus algorithms needed in blockchain.
As an example, in Input/Output Automata~\cite{Lyn03}, the number of specified states accessible by an asynchronous algorithm before the threshold is reached could be proportional to the number of permutations of message receptions. Executing the automated verification of an invariant 
could require a computation proportional to the number of these permutations. More dramatically, the Byzantine fault model typically allows some processes to send arbitrarily formed and arbitrarily many messages---making the number of states to explore potentially infinite. 
As a result, this is only with the recent definition of threshold automata reducing this state space
that we were able to verify our blockchain consensus components.

The remainder of the paper is organized as follows. Section~\ref{sec:vulnerabilities} presents new and existing problems affecting known blockchain Byzantine consensus. In Section~\ref{sec:method}, we explain how we verified a Byzantine fault tolerant broadcast abstraction common to multiple blockchains.
Section~\ref{sec:rw} presents the related work and Section~\ref{sec:conclusion} discusses our verifications and concludes the paper.
In Appendix~\ref{app:bbc}, we list the pseudocode, specification and verification experiments of a variant of the Byzantine consensus used in Red Belly Blockchain.

\section{The Problem of Proving Blockchain Consensus Algorithms by Hand}\label{sec:vulnerabilities}

In this section, we illustrate the risk of trying to prove blockchain consensus algorithms by hand by describing a list 
of safety and liveness limitations affecting the %
Byzantine fault tolerant
algorithms implemented in actual blockchain systems. These limitations, depicted in Table~\ref{table:problems}, are not necessarily errors in the proofs but stem 
from the ambiguous descriptions in prose rather than formal statements and the lack of machine-checked proofs.
As far as we know, until now no Byzantine fault tolerant consensus algorithms used in a blockchain had been formally verified automatically.

\begin{table*}
\caption{Consensus algorithms that experienced liveness or safety limitations\label{table:problems}}
\begin{center}
\setlength{\tabcolsep}{6pt}
\begin{tabular}{rllccl}
\hline
\cellcolor[gray]{0.75} Algorithms & \cellcolor[gray]{0.75} Ref. & \cellcolor[gray]{0.75} Limitation & \cellcolor[gray]{0.75}Counter-example & \cellcolor[gray]{0.75} Alternative & \cellcolor[gray]{0.75} Blockchain \\ \toprule
Randomized consensus & \cite{MMR14} & liveness & [new] &  \cite{MMR15}  & HoneyBadger~\cite{MXC16} \\
Casper & \cite{BG19} & liveness & [new] &\cite{ZRA18} & Ethereum v2.0~\cite{Eth19} \\
Ripple consensus & \cite{SYB14} & safety & \cite{AKM15} & \cite{CM18} & xRapid~\cite{Bro19} \\
Tendermint consensus & \cite{BKM18} & safety & \cite{ADB19} & \cite{ADP18b} & Tendermint~\cite{Kwo14} \\
Zyzzyva & \cite{KADCW09} & safety & \cite{AGD17} & \cite{AGZ15} & SBFT~\cite{GAGM18} \\
IBFT & \cite{Lin19} & liveness & \cite{Sal19} & \cite{Sal19} & Quorum~\cite{JPM18} \\
\bottomrule
\end{tabular}
\end{center}
\end{table*}

\subsection{The HoneyBadger and its randomized binary consensus}
HoneyBadger~\cite{MXC16} builds upon the combination of three algorithms from the literature to solve the Byzantine consensus with high probability in an asynchronous model. This protocol is being integrated in one of the most popular blockchain software, called Ethereum \texttt{parity}.\footnote{\url{https://forum.poa.network/t/posdao-white-paper/2208}.}
First it uses a classic reduction from the problem of multi-value Byzantine consensus to the problem of binary Byzantine consensus working in the asynchronous model. Second, it reuses a randomized Byzantine binary consensus algorithm~\cite{MMR14} that 
aims at terminating in expected constant time by using a common coin that returns the same unpredictable value at every process.   Third, it uses a common coin implemented with a threshold signature scheme~\cite{CKS00} that requires the participation of correct processes to return a value. 
\\

\noindent
{\bf Randomized binary consensus.}
In each asynchronous round of this randomized consensus~\cite{MMR14}, the processes ``binary value broadcast''---or ``BV-broadcast'' for short---their input binary value.
The binary value broadcast (detailed later in Section~\ref{sec:bvb}) simply consists of broadcasting (including to oneself) a value, then rebroadcasting (or \emph{echoing}) any value received from $t+1$ distinct processes and finally bv-delivering any value received from $2t+1$ distinct processes.
These delivered values are then broadcast to the other processes and all correct processes record,  into the set $\ms{values}$, the values received from $n-t$ distinct processes that are among the ones previously delivered. For any correct process $p$, if $\ms{values}$ happens to contain only the value $c$ returned by the common coin then $p$ decides this value, if $\ms{values}$ contains only the other binary value $\neg c$, then $p$ sets its estimate to this value and if $\ms{values}$ contains two values, then $p$ sets its estimate to $c$. Then $p$ moves to the next round until it decides.\\

\noindent
{\bf Liveness issue.}
The problem is that in practice, as the communication is asynchronous, the common coin cannot return at the exact same time at all processes. In particular, if some correct processes are still at the beginning of their round $r$ while the adversary observes the outcome of the common coin for round $r$ then the adversary can prevent progress among the correct processes by controlling messages between correct processes and by sending specific values to them. Even if a correct process invokes the common coin before the Byzantine process, then the Byzantine can prevent correct processes from progressing. \\

\noindent
{\bf Counter example.}
To illustrate the issue, we consider a simple counter-example with $n=4$ processes and $t=1$ Byzantine process.
Let $p_1$, $p_2$ and $p_3$ be correct processes with input values $0, 1, 1$, respectively, and let $p_4$ be a Byzantine process. 
The goal is for process $p_4$ to force some correct processes to deliver $\{0,1\}$ and another correct process to deliver $\{\neg c\}$ where $c$ is the value returned by the common coin in the current round.
As the Byzantine process has control over the network, it prevents $p_2$ from receiving anything 
before guaranteeing that $p_1$ and $p_3$ deliver $\{0,1\}$. 
It is easy to see that $p_4$ can force $p_1$ and $p_3$ to bv-deliver $1$ so let us see how $p_4$ forces $p_1$ and $p_3$ to deliver $0$.
Process $p_4$ sends $0$ to $p_3$ so that $p_3$ receives value $0$ from both $p_1$ and $p_4$, and thus echoes $0$. Then $p_4$ sends $0$ to $p_1$. Process $p_1$ then receives value $0$ from $p_3$, $p_4$ and itself, hence $p_1$ echoes and delivers $0$. Similarly, $p_3$
 receives value $0$ from $p_1$, $p_4$ and itself, hence $p_3$ delivers $0$.
 To conclude $p_1$ and $p_3$ deliver $\{0,1\}$.
Processes $p_1$, $p_3$ and $p_4$ invoke the coin and there are two cases to consider depending on the value returned by the coin $c$.
\begin{itemize}
\item {\bf Case $c=0$:} 
Process $p_2$ receives now 1 from $p_3$, $p_4$ and itself, so it delivers 1.
\item {\bf Case $c=1$:} 
This is the most interesting case, as $p_4$ should prevent some correct process, say $p_2$, from delivering $1$ even though $1$ is the most represented input value among correct processes.
Process $p_4$ sends $0$ to $p_2$  and $p_3$ so that both $p_2$ and $p_3$ receive value $0$ from $p_1$ and $p_4$ and thus both echo $0$. Due to $p_3$'s echo, $p_2$ receives $2t+1$ $0$s and $p_2$ delivers $0$.
\end{itemize}
At least two correct processes obtain $\ms{values} = \{0,1\}$ and another correct process can obtain $\ms{values} = \{\neg c\}$. It follows that the correct processes with $\ms{values} = \{0,1\}$ adopt $c$ as their new estimate while the correct process with $\ms{values} = \{\neg c\}$ takes $\neg c$ as its new estimate and no progress can be made within this round. Finally, if the adversary (controlling $p_4$ in this example) keeps this strategy, then it will produce an infinite execution without termination. \\

\noindent
{\bf Alternative and counter-measure.}
The problem would be fixed if we could ensure that the common coin always return at the correct processes before returning at a Byzantine process, however, we cannot distinguish a correct process from a Byzantine process that acted correctly.
We are thankful to the authors 
of the randomized algorithm for confirming our counter-example, they also wrote a remark in~\cite{MMR15} indicating that both a fair scheduler and a perfect common coin were actually needed for the consensus of~\cite{MMR14} to converge with high probability, however, no counter example motivating the need for a fair scheduler was proposed. 
The intuition behind the fair scheduler is that it requires to have the same probability of receiving messages in any order~\cite{BT85} and thus limits the power of the adversary on the network.
A new algorithm~\cite{MMR15} does not suffer from the same problem and offers the same  asymptotic complexity in message and time as~\cite{MMR14} but requires more communication steps, it could be used as an alternative randomized consensus in HoneyBadger to cope with this issue.

\subsection{The Ethereum blockchain and its upcoming Casper consensus}
Casper~\cite{ZRA18,BG19} is an alternative to the existing longest branch technique to agree on a common block within Ethereum. 
It is well-known that Ethereum can experience disagreement when different processes receive distinct blocks for the same index. These disagreements are typically resolved by waiting until the longest branch is unanimously identified. Casper aims at solving this issue by offering consensus. \\

\noindent
{\bf The Casper FFG consensus algorithm.}
The FFG variant of Casper is intended to be integrated to Ethereum v2.0 during phase~0~\cite{Eth19}.
It is claimed to ensure finality~\cite{BG19}, a property that may seem, at first glance, to result from the termination of consensus. %
The model of Casper assumes authentication, synchrony
and that strictly less than $1/3$ stake is owned by Byzantine processes. 
Casper builds a ``blockchain tree'' consisting of a partially ordered set of blocks. The genesis block as well blocks at indices multiple of 100 are called \emph{checkpoints}. 
Validator processes vote for a link between checkpoints of a common branch and a checkpoint is 
\emph{justified} if it is the initial, so-called \emph{genesis}, block or there is a link from a justified checkpoint pointing to it voted by a supermajority of $\lfloor \frac{2n}{3} \rfloor + 1$ validators. \\

\noindent
{\bf Liveness issue.}
Note first that Casper executes speculatively and that there is not a single consensus instance per level of the Casper blockchain tree. Each time an agreement attempt at some level of the tree fails due to the lack of votes for the same checkpoint, the height of the tree grows.
Unfortunately, it has been observed that nothing guarantees the termination of Casper FFG~\cite{CGG19} and we present below an example of infinite execution. \\

\noindent
{\bf Counter example.}
To illustrate why the consensus does not terminate in this model, 
let $h$ be the level of the highest block that is justified.
\begin{enumerate}
\item Validators try to agree on a block at level $h+k$ ($k>0$) by trying to gather $\lfloor \frac{2n}{3} \rfloor +1$ votes for the same block at level $h+k$ (or more precisely the same link from level $h$ to $h+k$). This may fail if, for example, $\frac{n}{3}$ validators vote for one of three distinct blocks at this level $h+k$.
\item Upon failure to reach consensus at level $h+k$, the correct validators, who have voted for some link from height $h$ to $h+k$ and are incentivised to abstain from voting on another link from $h$ to $h+k$, can now try to agree on a block at level $h+k'$ ($k'>k$), but again no termination is guaranteed.
\end{enumerate}
The same steps (1) and (2) may repeat infinitely often. Note that plausible liveness~\cite[Theorem 2]{BG19} is still fulfilled in that the supermajority `can' always be produced as long as you have infinite memory, but no such supermajority link is ever produced in this infinite execution.\\

\noindent
{\bf Alternative and counter-measure.}
Another version of Casper, called CBC, has also been proposed~\cite{ZRA18}. It is claimed to be ``correct by construction'', hence the name CBC. 
This could potentially be used as a replacement to FFG Casper for Ethereum v2.0 even in phase~0 for applications that require consensus, and thus termination.

\subsection{Known problems in blockchain Byzantine consensus algorithms}

To show that our two counter examples presented above are not isolated cases in the context of blockchains, we also 
list below four counter examples from the literature that were reported by colleagues and affect 
the Ripple consensus algorithm, Tendermint and Zyzzyva. This adds to the severity of the problem of proving algorithm by hand before using them in critical applications like blockchains. \\

\noindent
{\bf The XRP ledger and the quorums of the Ripple consensus.}
The Ripple consensus~\cite{SYB14} is  a consensus algorithm originally intended to be used in the blockchain system developed by the company Ripple. 
The algorithm is 
presented at a high level as an algorithm that uses unique node lists as a set of \emph{quorums} or mutually intersecting sets that each individual process must contact to guarantee that its request will be stored by the system or that it can retrieve consistent information about asset ownership. The original   but deprecated white paper~\cite{SYB14} assumed that quorums overlap by about 20\%.

Later, some researchers published an article~\cite{AKM15} indicating that the algorithm was inconsistent and 
listing the environmental conditions under which consensus would not be solved and its safety would be violated.
They offered a fix in order to remedy this inconsistency through the use of different assumptions, requiring that quorums overlap by stricly more than 40\%.
Finally, the Ripple consensus algorithm has been replaced by the XRP ledger consensus protocol~\cite{CM18} called ABC-Censorship-Resilience under synchrony in part to fix this problem.\\

\noindent
{\bf The Tendermint blockchain and its locking variant to PBFT.}
Tendermint~\cite{Kwo14} has similar phases as PBFT~\cite{CL02} and works with asynchronous rounds~\cite{DLS88}. In each round, processes propose values in turn (phase 1), the proposed value is prevoted (phase 2), precommitted when prevoted by sufficiently many\footnote{`Sufficiently many' processes stands for at least $\lfloor \frac{2n}{3} \rfloor +1$ among $n$ processes.} processes (phase 3) and decided when precommitted by sufficiently many processes. To progress despite failures, processes stay in a phase only for up to a timeout period.
A difference with PBFT is that a correct process produces a proof-of-lock of $v$ at round $r$ if it precommits $v$ at round $r$. A correct process can only prevote $v'$ if it did not precommit a conflicting value $v\neq v'$.

As we restate here, there exists a counter-example~\cite{ADP18b} that illustrates the safety issue
with four processes $p_1, p_2, p_3$ and $p_4$ among which $p_4$ is Byzantine that propose in the round of their index number.
In the first round, correct processes prevote $v$,  $p_1$ and $p_2$ lock $v$ in this round and precommit it, 
$p_1$ decides $v$ while $p_2$ and $p_3$ do not decide, before $p_1$ becomes slow.
In the second round, process $p_4$ informs $p_3$ that it prevotes $v$ so that 
$p_3$ prevotes, precommits and locks $v$ in round 2. 
In the third round, $p_3$ proposes $v$ locked in round 2, forcing $p_2$ to unlock $v$ 
and in the fourth round, $p_4$ forces $p_3$ to unlock $v$ in a similar way.
Finally, $p_1$ does not propose anything and $p_2$ proposes another value $v' \neq v$ that gets decided by all. It follows that correct processes $p_1$ and $p_2$ decide differently, which violates agreement. Since this discovery, Tendermint kept evolving and the authors of the counter example acknowledged that some of the issues they reported were fixed~\cite{ADB19}, the authors also informed us that they notified the developers but ignore whether this particular safety issue has been fixed. \\

\noindent
{\bf Zyzzyva and the SBFT concurrent fast and regular paths.}
Zyzzyva~\cite{KADCW09} is a Byzantine consensus that requires view-change and combines a fast path where a client can learn the outcome of the consensus in 3 message delays  and a regular path where the client needs to collect a commit-certificate with $2f+1$ responses where $f$ is the actual number of Byzantine faults. The same optimization is currently implemented in the SBFT permissioned blockchain~\cite{GAGM18} to speed up termination when all participants are correct and the communication is synchronous.

There exist counter-examples~\cite{AGD17} that illustrate how the safety property of Zyzzyva can be violated. The idea of one counter-example consists of creating a commit-certificate for a value $v$, then experiencing a first view-change (due to delayed messages) and deciding another value $v'$ for a given index before finally experiencing a second view-change that leads to undoing the former decision $v'$ but instead deciding $v$ at the same index. SBFT is likely to be immune to this issue as the counter example was identified by some of the authors of SBFT. But a simple way to cope with this issue is to prevent the two paths from running concurrently as in the simpler variant of Zyzzyva called Azyzzva~\cite{AGZ15}. \\

\noindent
{\bf The Quorum blockchain and its IBFT consensus.}
IBFT~\cite{Lin19} is a Byzantine fault tolerant consensus algorithm at the heart of the Quorum blockchain designed by Morgan Stanley. It is similar to PBFT~\cite{CL02} except that is offers a simplified version of the PBFT view-change by getting rid of new-view messages. 
It aims at solving consensus under partial synchrony.
The protocol assumes that no more than $t<n/3$ processes---usually referred by IBFT as ``validators''---are Byzantine.

As reported in~\cite{Sal19}, IBFT does not terminate in a partially synchronous network even when failures are crashes. 
More precisely IBFT cannot guarantee that if at least one honest validator is eventually able to produce a valid finalized block then the transaction it contains will eventually be added to the local transaction ledger of any other correct process.
IBFT~v2.x~\cite{Sal19} fixes this problem but requires a transaction to be submitted to all correct validators for this transaction to be eventually included in the distributed permissioned transaction ledger. The proof was made by hand and we are not aware of any automated proof of this protocol as of today. \\

\section{Methodology for Verifying Blockchain Components}\label{sec:method}

In this section, we explain how we  %
verified the binary value broadcast blockchain component using the Byzantine model checker without being experts in verification.\footnote{Although we are not experts in verification, we are thankful to  verification experts Igor Konnov and Josef Widder for discussions on the syntax of threshold automata and for confirming that our consensus agreement property was verified by ByMC when our initial runs were taking longer than expected.} Then we explain how this helped us verify the correctness of a variant of the binary consensus of DBFT used in Red Belly Blockchain. \\

\subsection{Preliminaries on ByMC and BV-broadcast}

\noindent
{\bf Byzantine model checker.}
Fault tolerant distributed algorithms, like the Byzantine fault tolerant broadcast primitive presented below, are often based on parameters, like the number $n$ of processes, the maximum number of Byzantine faults $t$ or the number of Byzantine faults $f$.
Threshold-guarded algorithms~\cite{KVW15,KLVW17} use these parameters to define threshold-based guard conditions that enable transitions to different states. 
Once a correct process receives a number of messages that reaches the threshold, it progresses by taking some transition to a new state.
To circumvent the undecidability of model checking on infinite systems, Konnov, Schmid, Veith and Widder introduce two parametric interval abstractions~\cite{KSVW13} that model 
\emph{(i)}~each process with a finite-state machine independent of the parameters and \emph{(ii)}~the whole system with abstract counters that quantify the number of processes in each state in order to obtain a finite-state system.
Finally, they group a potentially infinite number of runs into an execution schema in order to allow bounded model checking, based on an SMT solver, over all the possible execution schemas~\cite{KVW15}.
ByMC~\cite{KW18} verifies threshold automata with this model checking and has been used to prove various distributed algorithms, like atomic commit or reliable broadcast. Given a set of safety and liveness properties, it outputs traces showing that the properties are satisfied in all the reachable states of the threshold automaton. 
Until 2018, correctness properties were only verified on one round but more recently the threshold automata framework was extended to randomized algorithms, making possible to verify algorithms such as Ben-Or's randomized consensus under round-rigid adversaries~\cite{BKL19}.\\

\begin{algorithm*}[t]
	\caption{The binary value broadcast algorithm\label{algo:bv-bcast}}
	{\footnotesize
	\begin{algorithmic}[1]
		\Part{$\lit{bv-broadcast}(\lit{MSG}, \ms{val}, \ms{conts}, i)$} { \Comment{binary value broadcast filters out values proposed only by Byzantine proc.} \label{line:bvbcast-start}
			\State $\lit{broadcast}(\lit{BV}, \tup{val, i})$ \Comment{broadcast binary value $\ms{val}$} \label{line:bcast-1}
			\Repeat \Comment{re-broadcast a received value only if it is sufficiently represented}
				\If{$(\lit{BV}, \tup{v, *})$ received from $(t + 1)$ distinct processes but not yet broadcast} \label{line:bcast-2} \Comment{received from at least one correct}
					\State $\lit{broadcast}(\lit{BV}, \tup{v, i})$ \Comment{echo $v$} \label{line:bvbcast-rebcast} 							\EndIf
				\If{$(\lit{BV}, \tup{v, *})$ received from $(2t + 1)$ distinct processes} \Comment{received from a majority of correct}
					\State $\ms{conts} \leftarrow \ms{conts} \cup \{v\}$\label{line:BYZ-safe-14}\Comment{deliver $v$}%
				\EndIf%
			\EndRepeat \label{line:bvbcast-end}%
		}\EndPart%
	\end{algorithmic}%
	}
\end{algorithm*}%

\noindent
{\bf Binary value broadcast.}\label{sec:bvb}
The binary value broadcast~\cite{MMR14}, also denoted BV-broadcast, is a Byzantine fault tolerant communication abstraction used in blockchains~\cite{MXC16,CNG18} that works in an asynchronous network with reliable channels where the maximum number of Byzantine failures is $t < n/3$.
The BV-broadcast guarantees that no values broadcast exclusively by Byzantine processes can be delivered by correct processes. This 
helps limiting the power of the adversary to make sure that a Byzantine consensus algorithm converges towards a value. In particular, by requiring that all correct processes BV-broadcast their proposals, one can guarantee that all correct processes will eventually observe their proposals, regardless of the values proposed by Byzantine processes. 
The binary value broadcast finds applications in blockchains: First, it is implemented in HoneyBadger~\cite{MXC16} to detect that correct processes have proposed diverging values in order to toss a common coin, that returns the same result across distributed correct processes, to make them converge to a common decision. Second, Red Belly Blockchain~\cite{CNG18} and the accountable blockchain that derives from it~\cite{CGG19} implement the BV-broadcast to detect whether the protocol can converge towards the parity of the round number by simply checking that it corresponds to one of the values that were ``bv-delivered''.

The BV-broadcast abstraction satisfies the four following properties:
\begin{enumerate}%
\item BV-Obligation. 
If at least $(t+1)$ correct  processes BV-broadcast the same value $v$, $v$  
is eventually added to the set $\ms{conts}_i$ of each correct process $p_i$. 
\item BV-Justification. 
If $p_i$ is correct and  $v\in \ms{conts}_i$,  $v$ has been 
BV-broadcast by some correct process.  (Identification following from receiving more than $t$ 0s or 1s.)
\item BV-Uniformity. 
If a value $v$ is added to the set $\ms{conts}_i$ of a correct process $p_i$, 
eventually  $v\in \ms{conts}_j$ at every correct process $p_j$. 
\item BV-Termination. 
Eventually the set $\ms{conts}_i$ of  each correct process $p_i$  is not
empty.  
\end{enumerate}

\algnewcommand{\LeftComment}[1]{\Statex \(\triangleright\) #1}

\subsection{Automated verification of a blockchain Byzantine broadcast}

In this section, we describe how we used threshold automaton to specify the binary value broadcast algorithm and ByMC in order to verify the protocol automatically.  We recall the BV-broadcast algorithm as depicted in Algorithm~\ref{algo:bv-bcast}. The algorithm consists of having at least $n-t$ correct processes broadcasting a binary value. Once a correct process receives a value from $t+1$ distinct processes, it broadcasts it if it did not do it already. Once a correct process receives a value from $2t+1$ distinct processes, it delivers it. Here the delivery is modeled by adding the value to the set $\ms{conts}$, which will simplify the description of our variant of DBFT binary consensus in Appendix~\ref{app:bbc}. 
\\

\begin{figure*}[ht!]
\begin{center}
\includegraphics[scale=0.55]{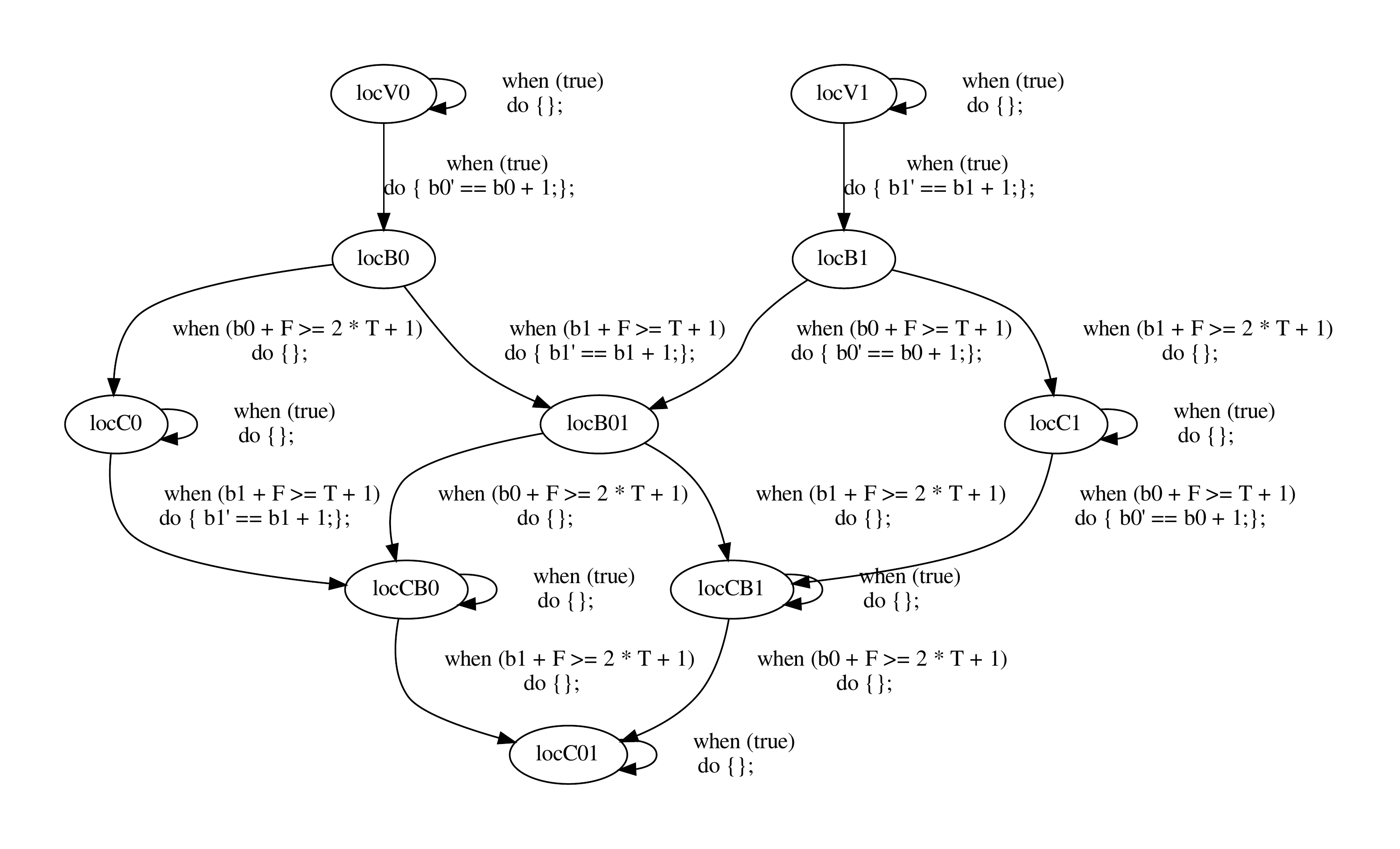}
\caption{The threshold automaton of the binary value broadcast algorithm\label{fig:graph}}
\label{tabvb}
\end{center}
\end{figure*}

\noindent
{\bf Specifying the distributed algorithm in a threshold automaton.}
Let us describe how we specify Algorithm~\ref{algo:bv-bcast} as a threshold automaton depicted in Figure~\ref{fig:graph}. 
Each state of the automaton or node in the corresponding graph represents a local state of a process. A process can move from one state to another thanks to an edge, called a \emph{rule}. A rule has the form $\phi \mapsto u$, where $\phi$ is a guard and $u$ an action on the shared variables. When the guard evaluates to true (e.g., more than $t+1$ messages of a certain type have been sent), the action is executed (e.g., the shared variable $s$ is incremented).

In Algorithm~\ref{algo:bv-bcast}, we can see that only two types of messages are exchanged: process $i$ can only send either $(\lit{BV}, \langle 0, i \rangle)$ or $(\lit{BV}, \langle 1, i \rangle)$. Each time a value is sent by a correct process, it is actually broadcast to all processes. Thus, we only need two shared variables $b0$ and $b1$ corresponding to the value $0$ and $1$ in the automaton (cf. Figure~\ref{fig:graph}). Incrementing $b0$ is equivalent to broadcasting $(\lit{BV}, \langle 0, i \rangle)$.
Initially, each correct process immediately broadcasts its value.  This is why the guard for the first rule is $\lit{true}$: a process in $locV0$ can immediately move to $locB0$ and send $0$ during the transition.

We then enter the \textit{repeat} loop of the pseudocode. The two \textit{if} statements are easily understandable as threshold guards. If more than $t+1$ messages with value $1$ are \textit{received}, then the process should broadcast $1$ (i.e., incremenent $b1$) since it has not already been done. Interestingly, the corresponding guard is $b1 + f \geq t + 1$. Indeed, the shared variable $b1$ only counts the messages \textit{sent} by correct processes. However, the $f$ faulty processes might send messages with arbitrary values. We want to consider all the possible executions, so the earliest moment a correct process can move from $locB0$ to $locB01$ is when the $f$ faulty processes and $t+1-f$ correct processes have sent $1$. The other edge leaving $locB0$ corresponds to the second \textit{if} statement, that is satisfied when $2t+1$ messages with value $0$ have been received. In state $locC0$, the value $0$ has been delivered. A process might stay in this state forever, so we add a self-loop with guard condition set to \textit{true}. 

After the state $locC0$, a process is still able to broadcast $1$ and eventually deliver $1$ after that. After the state $locB01$, a process is able to deliver $0$ and then deliver $1$, or deliver $1$ first and then deliver $0$, depending on the order in which the guards are satisfied. 
Apart from the self-loops, we remark that the automaton is a directed acyclic graph. On every path of the graph, we can verify that a shared variable is incremented only once. This is because in the pseudocode, a value can be broadcast only if it has not been broadcast before.\\
Finally, the states of the automaton correspond to the following (unique) situations for a correct process:

\begin{itemize}
	\item \textbf{locV0.} Initial state with value $0$, nothing has been broadcast nor delivered
	\item \textbf{locV1.} Initial state with value $1$, nothing has been broadcast nor delivered
	\item \textbf{locB0.} Only $0$ has been broadcast, nothing has been delivered
	\item \textbf{locB1.} Only $1$ has been broadcast, nothing has been delivered
	\item \textbf{locB01.}  Both $0$ and $1$ have been broadcast, nothing has been delivered 
	\item \textbf{locC0.} Only $0$ has been broadcast, only $0$ has been delivered
	\item \textbf{locCB0.}  Both $0$ and $1$ have been broadcast, only $0$ has been delivered
	\item \textbf{locC1.} Only $1$ has been broadcast, only $1$ has been delivered
	\item \textbf{locCB1.}  Both $0$ and $1$ have been broadcast, only $1$ has been delivered
	\item \textbf{locC01.}  Both $0$ and $1$ have been broadcast, both $0$ and $1$ have been delivered
\end{itemize}

Once the pseudocode is converted into a threshold automaton depicted in Figure~\ref{fig:graph}, one can simply write the corresponding specification in the threshold automata language to obtain the specification listed below (Figure \ref{app:ta}) for completeness. \\

\noindent
{\bf Defining the correctness properties and fairness assumptions.}
The above automaton is only the first half of the verification work. The second half consists in specifying the correctness properties that we would like to verify on the algorithm. We use temporal logic on the algorithm variables (number of processes in each location, number of messages sent and parameters) to formalize the properties. 
In the case of the BV-broadcast, the BV-Justification property of the BV-broadcast is: ``If $p_i$ is correct and  $v\in \ms{conts}_i$,  $v$ has been 
BV-broadcast by some correct process''. Given $\Diamond$, $\rightarrow$ and $||$ with the LTL semantics of `eventually', `implies' and `or', respectively, we translate this property in two specifications:

\begin{equation*}
  \left\{
  \begin{array}{cc}
  \ms{justification0}: & (\Diamond(\ms{locC0} \neq 0 ~||~ \ms{locC01} \neq 0)) \rightarrow \\
  ~~~~(\ms{locV0} \neq 0) \\
  \ms{justification1}: & (\Diamond(\ms{locC1} \neq 0 ~||~ \ms{locC01} \neq 0)) \rightarrow \\
  ~~~~(\ms{locV1} \neq 0)
  \end{array}
  \right.
\end{equation*}

Liveness properties are longer to specify, because we need to take into account some fairness constraints. Indeed, a threshold automaton describes processes evolving in an asynchronous setting without additional assumptions. An execution in which a process stays in a state forever is a valid execution, but it does not make any progress. 
If we want to verify some liveness properties, we have to add some assumptions in the specification. 
For instance, we can require that processes eventually leave the states of the automaton as long as they have received enough messages to enable the condition guarding the outgoing rule. In other words, a liveness property will be specified as:

$$\ms{liveness\_property}: \ms{fairness\_condition} \rightarrow \ms{property}$$

Note that this assumption is natural and differs from the round rigidity assumption that requires the adversary to eventually take any applicable transition of an infinite execution.
Finally, we wrote a threshold automaton specification whose \texttt{.ta} file is presented in Figure~\ref{app:ta} in only 116 lines.\\

\begin{figure*}
\begin{lstlisting}[showstringspaces=false,
stepnumber=1,
numbers=left,
tabsize=2,
basicstyle=\footnotesize, 
%
multicols=2,
  columns=flexible,
  basicstyle=\footnotesize\ttfamily
  ={\small\ttfamily}]
thresholdAutomaton Proc {
 local pc; shared b0, b1; 
 parameters N, T, F; 

 assumptions (0) { N>3*T; T>=F; T>=1; }

 locations (0) {
  locV0:[0]; locV1:[1]; locB0:[2]; 
  locB1:[3]; locB01:[4]; locC0:[5]; 
  locC1:[6]; locCB0:[7]; 
  locCB1:[8]; locC01:[9]; 
 }

 inits (0) {
  (locV0+locV1)==N-F;
  locB0==0; locB1==0; locB01==0; 
  locC0==0; locC1==0; locCB0==0;
  locCB1==0; locC01==0; b0==0; b1==0;
 }

 rules (0) {
% for v in [0, 1]:
 1: locV${v} -> locB${v}
   when (true)
   do { b${v}'==b${v}+1;
      unchanged(b${1-v}); };

 2: locB${v} -> locB01
   when (b${1-v}+F>=T+1)
   do { b${1-v}'==b${1-v}+1;
      unchanged(b${v}); };

 3: locB${v} -> locC${v}
   when (b${v}+F>=2*T+1)
   do { unchanged(b0, b1); };

 2: locC${v} -> locCB${v}
   when (b${1-v}+F>=T+1)
   do { b${1-v}'==b${1-v}+1;
      unchanged(b${v}); };

 3: locB01 -> locCB${v}
   when (b${v}+F>=2*T+1)
   do { unchanged(b0, b1); };

 3: locCB${v} -> locC01
   when (b${1-v}+F>=2*T+1)
   do { unchanged(b0, b1); };

 /* self loops */
 10: locV${v} -> locV${v}
   when (true) do {unchanged(b0, b1);};

 10: locC${v} -> locC${v}
   when (true) do {unchanged(b0, b1);};

 10: locCB${v} -> locCB${v}
   when (true) do {unchanged(b0, b1);};
% endfor

 10: locC01 -> locC01
   when (true) do {unchanged(b0, b1);};
 }

 specifications (0) {
  
 % for v in [0,1]:
  obligation${v}: 
   <>[]((locV0==0) && (locV1==0) &&
(locB0==0 || b1<T+1) && (locB1==0 || b0<T+1) &&
(locB0==0 || b0<2*T+1) && (locB1==0 || b1<2*T+1) &&
(locB01==0 || b0<2*T+1) && (locB01==0 || b1<2*T+1) &&
(locC0==0 || b1<T+1) && (locC1==0 || b0<T+1) &&
(locCB0==0 || b1<2*T+1) && (locCB1==0 || b0<2*T+1))
    ->
    ((locV${v}>=T+1)
     ->
     <>(locV0==0 && locV1==0 &&
      locB0==0 && locB1==0 &&
      locB01==0 && locC${1-v}==0 &&
      locCB${1-v}==0));

  justification${v}: (<>(locC${v}!=0 
  || locCB${v}!=0 || locC01!=0)) 
  -> (locV${v}!=0);

 uniformity${v}:
  <>[]((locV0==0) && (locV1==0) &&
(locB0==0 || b1<T+1) && (locB1==0 || b0<T+1) &&
(locB0==0 || b0<2*T+1) && (locB1==0 || b1<2*T+1) &&
(locB01==0 || b0<2*T+1) && (locB01==0 || b1<2*T+1) &&
(locC0==0 || b1<T+1) && (locC1==0 || b0<T+1) && 
(locCB0==0 || b1<2*T+1) && (locCB1==0 || b0<2*T+1))
		 ->
 (<>(locC${v}!=0 || locCB${v}!=0 || locC01!=0)
		  ->
  <>[](locC${1-v}==0 && locCB${1-v}==0));
 % endfor

 termination: 
  <>[]((locV0==0) && (locV1==0) &&
     (locB0==0 || b1<T+1) &&
     (locB1==0 || b0<T+1) &&
     (locB0==0 || b0<2*T+1) &&
     (locB1==0 || b1<2*T+1) &&
     (locB01==0 || b0<2*T+1) &&
     (locB01==0 || b1<2*T+1) &&
     (locC0==0 || b1<T+1) &&
     (locC1==0 || b0<T+1) && 
     (locCB0==0 || b1<2*T+1) &&
     (locCB1==0 || b0<2*T+1))
    ->
    <>(locV0 ==0 && locV1 ==0 &&
      locB0 ==0 && locB01==0);
  }
} /* Proc */
\end{lstlisting}

\caption{Threshold automaton specification for the binary value broadcast communication primitive}
\label{app:ta}
\end{figure*}

\noindent
{\bf Experimental results.} On a simple laptop with an Intel Core i5-7200U CPU running at 2.50GHz, verifying all the correctness properties for BV-broadcast takes less than 40 seconds. For simple properties on well-specified algorithms, such as the ones of the benchmarks included with ByMC, the verification time can be less than one second. This result encouraged us to verify a complete Byzantine consensus algorithm that builds upon the binary-value broadcast. \\

\noindent
{\bf Debugging the manual conversion of the algorithm to the automaton.} It is common that the specification does not hold at first try, because of some mistakes in the threshold automaton model or in the translation of the correctness property into a formal specification. In such cases, ByMC provides a detailed output and a counter-example showing where the property has been violated. We reproduced such a counter-example in Figure~\ref{cex} with an older preliminary version of our specification. This specification was wrong because a liveness property did not hold. ByMC gave parameters and provided an execution ending with a loop, such that the condition of the liveness was never met. This trace helped us understand the problem in our specification and allowed us to fix it to obtain the correct specification we illustrated before in Figure~\ref{app:ta}. 
Building upon this successful result, we specified a more complex Byzantine consensus algorithm
that uses the same broadcast abstraction but we did not encounter any bug during this process and our first specification was proved correct by ByMC.
By lack of space we defer its pseudocode, threshold automaton specification and experimental results in Appendix~\ref{app:bbc}.

\begin{figure*}

\begin{lstlisting}[language=Prolog,
	showstringspaces=false,
	stepnumber=1,
	numbers=left,
	tabsize=2,
	basicstyle=\footnotesize, 
	commentstyle=\color{gray},
	  basicstyle=\small\ttfamily
	  ={\small\ttfamily},breaklines=true]
N:=34; T:=11; F:=1;
0 (F 0) x 0: b0:=0; b1:=0; K[pc:0]:=21; K[pc:1]:=12; K[*]:=0;
1 (F 1) x 1: b0:=1; K[pc:0]:=20; K[pc:2]:=1;

                          (...)

24 (F 52) x 1:  b1:=21; K[pc:5]:=12; K[pc:7]:=21;
****************
b0:=33; b1:=21; K[pc:0]:=0; K[pc:1]:=0; K[pc:2]:=0; 
K[pc:3]:=0; K[pc:4]:=0; K[pc:5]:=12; K[pc:6]:=0; K[pc:7]:=21; 
K[pc:8]:=0; K[pc:9]:=0;

****** LOOP *******
N:=34; T:=11; F:=1;
25 (F 83) x 1:  <self-loop>
****************
K[pc:2]:=0; K[pc:4]:=0; K[pc:5]:=12; K[pc:7]:=21; K[pc:8]:=0; 
K[pc:9]:=0;
\end{lstlisting}
	
\caption{Truncated counter-example produced by ByMC for a faulty specification of BV-broadcast}
\label{cex}
\end{figure*}

\section{Related Work}\label{sec:rw}
The observations that some of the blockchain consensus proposals have issues is not new~\cite{Gra16,CV17}. It is now well known that the termination of existing blockchain like 
Ethereum requires an additional assumption like synchrony~\cite{Gra16}. 
Our Ethereum counter-example differs as it considers the upcoming consensus algorithm of Ethereum v2.0. In~\cite{CV17}, the conclusions  are different from ours as they generalize on other Byzantine consensus proposals, like Tangaroa, not necessarily in use in blockchain systems. Our focus is on  consensus used in blockchains that are trading valuable assets because these are critical applications.

Threshold automata already proved helpful to automate the proof of  existing consensus algorithms~\cite{KW18}. 
They have even been useful in illustrating why a specification of the King-Phase algorithm~\cite{BG89} was incorrect~\cite{SKWZ19} (due to the strictness of a lower symbol), later fixed in~\cite{BSW11}.
We did not list this as one of the inconsistency problems that affects blockchains 
as we are not aware of any blockchain implementation that builds upon the King-Phase algorithm. 
In~\cite{LWB17}, the authors use threshold guarded automata to prove two broadcast primitives and the Bosco Byzantine consensus correct, however, Bosco offers a fast path but requires another consensus algorithm for its fallback path so its correctness depends on the assumption that it relies on a correct consensus algorithm.

In general it is hard to formally prove algorithms that work in a partially synchronous model while there exist tools to reduce the state space of synchronous consensus to finite-state model checking~\cite{ARS18}. Part of the reason is that common partially synchronous solutions attempt to give sufficient time to processes in different asynchronous round by incrementing a timeout until the timeout is sufficiently large to match the unknown message delay bound. PSync~\cite{CHZ16} and ConsL~\cite{MSB17} are languages that help reasoning formally about partially synchronous algorithms. In particular, ConsL was shown effective at verifying consensus algorithms but only for the crash fault tolerant model. Here we used the ByMC model checker for asynchronous Byzantine fault tolerant systems and require the round-rigidity assumption to show a variant of the binary consensus of DBFT~\cite{CGLR18}.

A framework allows to build certified proofs of distributed algorithms with the proof assistant Coq~\cite{ACD16}. The tools developed in this framework are for the termination of self-stabilizing algorithms. It is unclear how it can be easily applied to complex algorithms like Byzantine consensus algorithms. 
Another model for distributed algorithms has been encoded in the interactive proof assistant Isabelle/HOL, and used to verify several consensus algorithms \cite{CBDM11}.

In~\cite{YPL99}, the authors present TLC, a model checker for debugging a finite-state model of a TLA+ specification. TLA+ is a specification language for concurrent and reactive systems that builds upon the temporal logic TLA. One limitation is that the TLA+ specification might comprise an infinite set of states for which the model checker can only give a partial proof. In order to run the TLC model checker on a TLA+ specification, it is necessary to fix the parameters such as the number of processes $n$ or the bounds on integer values. In practice, the complexity of model checking explodes rapidly and makes it difficult to check anything beyond toy examples with a handful of processes. TLC remains useful---in particular in industry---to prove that some specifications are wrong~\cite{N14}.
TLA+ also comes with a proof system called TLAPS. TLAPS supports manually written hierarchically structured proofs, which are then checked by backend engines such as Isabelle, Zenon or SMT solvers~\cite{CDLMRV12}. TLAPS is still being actively developed but it is already possible---albeit technical and lengthy---to prove algorithms such as Paxos.

\section{Discussion and Conclusion}\label{sec:conclusion}
In this paper, we argued for the formal verification of blockchain Byzantine fault tolerant algorithms as a way to reduce the numerous issues resulting from non-formal proofs for such critical applications as blockchains. 
In particular, we illustrated the problem with new counter-examples of algorithms at the core of widely deployed blockchain software. 

We show that it is now feasible, for non experts, to verify blockchain Byzantine components on modern machines thanks to the recent advances in formal verification and illustrate it with relatively simple specifications of a broadcast abstraction common to multiple blockchains as well as a variant of the Byzantine consensus algorithm of Red Belly Blockchain.

To verify the Byzantine consensus, we	%
assumed a round rigid adversary that schedules transitions in a fair way.
This is not new as in~\cite{BKL19} the model checking of the randomized algorithm from Ben-Or required a round-rigid adversary.
Interestingly, we do not need this assumption to verify the binary value broadcast abstraction that works in an asynchronous model.
As future work, we would like to prove other Byzantine fault tolerant algorithmic components of blockchain systems.

\subsection*{Acknowledgements}
We wish to thank Igor Konnov and Josef Widder for helping us understand the syntax and semantics of the threshold automata specification language and for confirming that ByMC verified the agreement1 property of our initial specification. We thank Tyler Crain, Achour Most\'{e}faoui and Michel Raynal for discussions of the HoneyBadger counter-example, and Yackolley Amoussou-Guenou, Maria Potop-Butucaru and Sara Tucci for discussions on the Tendermint counter-example. This research is supported under Australian Research Council Discovery Projects funding scheme (project number 180104030) entitled ``Taipan: A Blockchain with Democratic Consensus and Validated Contracts'' and Australian Research Council Future Fellowship funding scheme (project number 180100496) entitled ``The Red Belly Blockchain: A Scalable Blockchain for Internet of Things''.

\bibliographystyle{abbrv}
\bibliography{references}

\appendix
\section{Verifying a blockchain Byzantine consensus algorithm}\label{app:bbc}

The Democratic Byzantine Fault Tolerant consensus algorithm~\cite{CGLR18} is a Byzantine consensus algorithm that does not require a leader. It was implemented in Red Belly Blockchain~\cite{CNG18} to offer high performance through multiple proposers and was used in Polygraph~\cite{CGG19} to detect malicious participants responsible of disagreements when $t\geq n/3$.
As depicted in Algorithm~\ref{alg:bbc-safe},
its binary consensus proceeds in asynchronous rounds that correspond to the iterations of a loop where correct processes refine their estimate value.

\begin{algorithm*}[h!]
	\caption{A variant of DBFT binary Byzantine consensus algorithm\label{alg:bbc-safe}}
	{\footnotesize
	\begin{algorithmic}[1]

	\Statex{\textit{Notation: "Received $k$ messages" is a shortcut for "Received $k$ messages from different processes in the same round $r$ as the current round."}}
	\Statex
	\Part{$\lit{propose}(v)$}{
		\State $\ms{est} \gets v$ \Comment{initial estimate is the proposed value}
		\State $r \gets 0$ \Comment{initialize the round number}

		\Repeat \label{line:round-start} \Comment{repeat in asynchronous rounds}
			\State $r \leftarrow r + 1$; \label{line:inc-rnumber} \Comment{increment the round number}
			\State $\lit{broadcast}(\ms{tag} = \lit{BV}, \ms{round} = r, \ms{value} = \ms{est})$ \Comment{initial broadcast} 
			\While{$\lit{true}$} \Comment{start of binary value broadcast phase}
				\If{received $(t+1)$ $\lit{BV}$ messages with value $w$ and $w$ not broadcast yet}  \Comment{received from at least one correct}
				\State $\lit{broadcast}(\ms{tag} = \lit{BV}, \ms{round} = r, \ms{value} = w)$ \Comment{rebroadcast legitimate estimates}
				\EndIf

				\If{received $(2t+1)$ $\lit{BV}$ messages with value $w$} \Comment{received from a majority of correct}
				\State $\lit{broadcast}(\ms{tag} = \lit{ECHO}, \ms{round} = r, \ms{value} = w)$ \Comment{broadcast ECHO message}
				\State $\lit{break}$ \Comment{exit the while loop to proceed to next phase}
				\EndIf
				
			\EndWhile
			\While{$\lit{true}$} \Comment{wait to have received enough messages}
				\State $\ms{echoes} \gets \{w \in \{0,1\}: \text{received } (2t+1)$ $\lit{BV}\text{ messages with value }w\}$ \Comment{check the bv-delivered messages}
				\If{received $(n-t)$ $\lit{ECHO}$ messages with value $w \in \ms{echoes}$} \Comment{received singletons from sufficiently many}
				\State $\ms{est} \gets w$ \Comment{refine estimate} \label{line:refine-est}
				\If{$w = r \mod 2$ and not decided yet} \Comment{depending on the singleton value $w$...}
				\State $\lit{decide}(w)$ \Comment{...decide the parity of the round}
				\EndIf
				\State $\lit{break}$ \Comment{exit the while loop to proceed to next round}

				\EndIf
				\If{received $(n-t)$ $\lit{ECHO}$ messages and $\ms{echoes} = \{0,1\}$} \Comment{all values were bv-delivered}
				\State $\ms{est} \gets r \mod 2$  \Comment{set estimate to round parity}
				\State $\lit{break}$ \Comment{exit the while loop to proceed to next round}
				\EndIf

			\EndWhile
			\If{decided in round $r_i-2$} $\lit{exit}$ \Comment{exit the consensus only after having helped others decide}
			\EndIf
			
			\EndRepeat \label{line:round-end}
		}\EndPart

		\Statex

	\end{algorithmic}%
	}%
\end{algorithm*}%

\begin{figure*}[ht!]
	\begin{center}
	\includegraphics[scale=0.5]{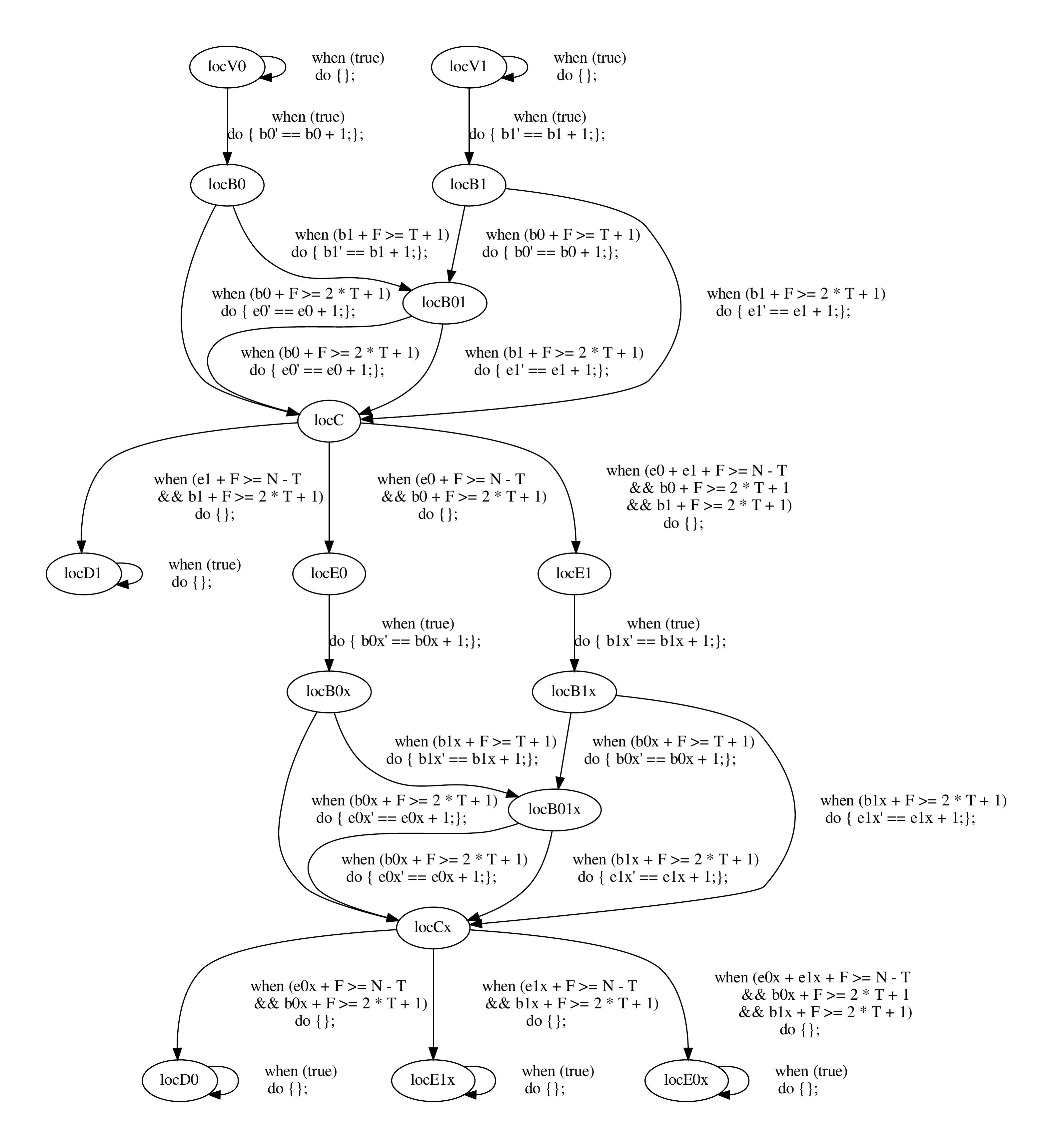}
	\caption{The threshold automaton of the DBFT binary consensus variant}
	\label{tadbft}
	\end{center}
	\end{figure*}

Initially, each correct process sets its estimate to its input value. 
Correct processes broadcast 
these estimate and rebroadcast only values received by $t+1$ distinct processes because they 
are proposed by correct processes.
Each value received from $2t+1$  distinct processes (and from a majority of correct processes) is stored in the \emph{echoes} set and is broadcast as part of an $\lit{ECHO}$ message. 
The $\lit{ECHO}$ value received from $n-t$ distinct processes that also belongs to $\ms{echoes}$ becomes the new estimate (line~\ref{line:refine-est}) for the next round.
If this value corresponds to the parity of the round, then the correct process decides this value.
If \emph{echoes} contains both values, then the estimate  for the next round becomes the parity of the round. 
As opposed to the original and partially synchronous deterministic version~\cite{CGLR18}, this variant 
uses one less broadcast phase and
offers termination in an asynchronous network under round-rigidity that requires the adversary to eventually perform any applicable transition within an infinite execution. This assumption was previously used to show termination of another algorithm with high probability~\cite{BKL19}. Below we show the specification of our consensus algorithm in threshold automata.

\begin{lstlisting}[language=Prolog,
showstringspaces=false,
stepnumber=1,
%
tabsize=2,
basicstyle=\footnotesize, 
commentstyle=\color{gray},
%
  columns=flexible,
  basicstyle=\small\ttfamily
  ={\small\ttfamily}]
/*
 Variant of the Safe DBFT binary Byzantine 
 consensus
 */
thresholdAutomaton Proc {

  local pc;

  /* Messages sent by correct processes */
  /* First round */
  shared b0, b1; 
  shared e0, e1; 
  /* Second round */
  shared b0x, b1x;
  shared e0x, e1x; 

  parameters N, T, F;   

  assumptions (0) {
    N > 3 * T;
    T >= F;
    T >= 1;
  }

  locations (0) {
    locV0:    [0]; 
    locV1:    [1];
    locB0:    [2]; 
    locB1:    [3]; 
    locB01:   [4];
    locC:     [5];
    locE0:    [6]; 
    locE1:    [7]; 
    locD1:    [8];
    locB0x:   [9]; 
    locB1x:   [10]; 
    locB01x:  [11];
    locCx:    [12];
    locE0x:   [13]; 
    locE1x:   [14]; 
    locD0:    [15];
  }

  inits (0) {
    (locV0 + locV1) == N - F;
    
    locB0   == 0;
    locB1   == 0;
    locB01  == 0;
    locC    == 0;
    locE0   == 0;
    locE1   == 0;
    locD1   == 0;
    locB0x  == 0;
    locB1x  == 0;
    locB01x == 0;
    locCx   == 0;
    locE0x  == 0;
    locE1x  == 0;
    locD0   == 0;

    b0  == 0;
    b1  == 0;
    e0  == 0;
    e1  == 0;
    b0x == 0;
    b1x == 0;
    e0x == 0;
    e1x == 0;
  }

  rules (0) {
% for v in [0, 1]:
  1: locV${v} -> locB${v}
      when (true)
      do { b${v}' == b${v} + 1;
           unchanged(b${1-v}, e0, e1);
           unchanged(b0x, b1x, e0x, e1x);
         };
% endfor

% for v in [0, 1]:
  2: locB${v} -> locB01
      when (b${1-v} + F >= T + 1)
      do { b${1-v}' == b${1-v} + 1;
           unchanged(b${v}, e0, e1);
           unchanged(b0x, b1x, e0x, e1x);
         };
% endfor

% for v in [0, 1]:
  3: locB${v} -> locC
      when (b${v} + F >= 2 * T + 1)
      do { e${v}' == e${v} + 1;
           unchanged(b0, b1, e${1-v});
           unchanged(b0x, b1x, e0x, e1x);
         };
% endfor

% for v in [0, 1]:
  4: locB01 -> locC
      when (b${v} + F >= 2 * T + 1)
      do { e${v}' == e${v} + 1;
           unchanged(b0, b1, e${1-v});
           unchanged(b0x, b1x, e0x, e1x);
         };
% endfor

  5: locC -> locD1
      when (e1 + F >= N - T
            && b1 + F >= 2 * T + 1)
      do {
           unchanged(b0, b1, e0, e1);
           unchanged(b0x, b1x, e0x, e1x);
      };

  6: locC -> locE0
      when (e0 + F >= N - T
            && b0 + F >= 2 * T + 1)
      do {
           unchanged(b0, b1, e0, e1);
           unchanged(b0x, b1x, e0x, e1x);
      };

  7: locC -> locE1
      when (e0 + e1 + F >= N - T
            && b0 + F >= 2 * T + 1
            && b1 + F >= 2 * T + 1)
      do {
           unchanged(b0, b1, e0, e1);
           unchanged(b0x, b1x, e0x, e1x);
      };

% for v in [0, 1]:
  8: locE${v} -> locB${v}x
      when (true)
      do { b${v}x' == b${v}x + 1;
           unchanged(b0, b1, e0, e1);
           unchanged(b${1-v}x, e0x, e1x);
         };
% endfor

% for v in [0, 1]:
  9: locB${v}x -> locB01x
      when (b${1-v}x + F >= T + 1)
      do { b${1-v}x' == b${1-v}x + 1;
           unchanged(b0, b1, e0, e1);
           unchanged(b${v}x, e0x, e1x);
         };
% endfor

% for v in [0, 1]:
  10: locB${v}x -> locCx
      when (b${v}x + F >= 2 * T + 1)
      do { e${v}x' == e${v}x + 1;
           unchanged(b0, b1, e0, e1);
           unchanged(b0x, b1x, e${1-v}x);
         };
% endfor

% for v in [0, 1]:
  11: locB01x -> locCx
      when (b${v}x + F >= 2 * T + 1)
      do { e${v}x' == e${v}x + 1;
           unchanged(b0, b1, e0, e1);
           unchanged(b0x, b1x, e${1-v}x);
         };
% endfor

  12: locCx -> locD0
      when (e0x + F >= N - T
            && b0x + F >= 2 * T + 1)
      do {
           unchanged(b0, b1, e0, e1);
           unchanged(b0x, b1x, e0x, e1x);
      };

  13: locCx -> locE1x
      when (e1x + F >= N - T
            && b1x + F >= 2 * T + 1)
      do {
           unchanged(b0, b1, e0, e1);
           unchanged(b0x, b1x, e0x, e1x);
      };

  14: locCx -> locE0x
      when (e0x + e1x + F >= N - T
            && b0x + F >= 2 * T + 1
            && b1x + F >= 2 * T + 1)
      do {
           unchanged(b0, b1, e0, e1);
           unchanged(b0x, b1x, e0x, e1x);
      };

  /* self loops */

% for v in [0, 1]:
  10: locV${v} -> locV${v}
      when (true)
      do {
          unchanged(b0, b1, e0, e1);
          unchanged(b0x, b1x, e0x, e1x);
      };
% endfor

% for v in [0, 1]:
  10: locD${v} -> locD${v}
      when (true)
      do {
          unchanged(b0, b1, e0, e1);
          unchanged(b0x, b1x, e0x, e1x);
      };
% endfor

% for v in [0, 1]:
  10: locE${v}x -> locE${v}x
      when (true)
      do {
          unchanged(b0, b1, e0, e1);
          unchanged(b0x, b1x, e0x, e1x);
      };
% endfor
  }
  specifications (0) {

  % for v in [0, 1]:
      validity${v}: 
      (locV${1-v} == 0) -> 
      [](locD${1-v} == 0 && locE${1-v}x == 0);
  % endfor

  % for v in [0, 1]:
    agreement${v}: 
    []((locD${v} != 0) -> 
    [](locD${1-v} == 0 && locE${1-v}x == 0));
  % endfor

    round_termination: 
      <>[](
        (locV0   == 0) &&
        (locV1   == 0) &&
        (locB0   == 0 || (b1 < T + 1 && b0 < 2 * T + 1)) &&
        (locB1   == 0 || (b0 < T + 1 && b1 < 2 * T + 1)) &&
        (locB01  == 0 || (b0 < 2 * T + 1 && b1 < 2 * T + 1)) &&
        (locC    == 0 || 
            ((e1 < N - T || b1 < 2 * T + 1) && 
            (e0 < N - T || b0 < 2 * T + 1) && 
            (e0 + e1 < N - T || 
                  b0 < 2 * T + 1 || 
                  b1 < 2 * T + 1) )) &&
        (locE0   == 0) &&
        (locE1   == 0) &&
        (locB0x  == 0 || (b1x < T + 1 && b0x < 2 * T + 1)) &&
        (locB1x  == 0 || (b0x < T + 1 && b1x < 2 * T + 1)) &&
        (locB01x == 0 || 
            (b0x < 2 * T + 1 && b1x < 2 * T + 1)) &&
        (locCx   == 0 || 
            ((e1x < N - T || b1x < 2 * T + 1) && 
            (e0x < N - T || b0x < 2 * T + 1) && 
            (e0x + e1x < N - T || 
                  b0x < 2 * T + 1 || 
                  b1x < 2*T+1)))
      )
      ->
      <>(
        locV0   == 0 &&
        locV1   == 0 &&
        locB0   == 0 &&
        locB1   == 0 &&
        locB01  == 0 &&
        locC    == 0 &&
        locE0   == 0 &&
        locE1   == 0 &&
        locB0x  == 0 &&
        locB1x  == 0 &&
        locB01x == 0 &&
        locCx   == 0
      );

  }
} /* Proc */
\end{lstlisting}

\begin{figure*}[h]
	\begin{center}
	\caption{Time to verify the Byzantine consensus of Algorithm~\ref{alg:bbc-safe}\label{fig:histogram}}
	\includegraphics[scale=0.8]{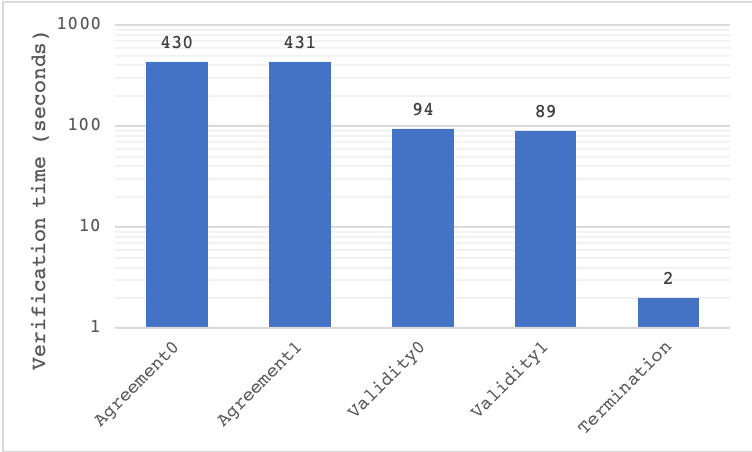}
	\end{center}
\end{figure*}

  \noindent
  {\bf Experimental results.} The Byzantine consensus algorithm has far more states and variables than the BV-broadcast primitive and it is too complex to be verified on a personnal computer. We ran the parallelized version of ByMC with MPI on a 4 AMD Opteron 6276 16-core CPU with 64 cores at 2300 MHz with 64\,GB of memory. %
The verification times for the 5 properties are listed in Figure~\ref{fig:histogram} and sum up to 1046 seconds or 17 minutes and 26 seconds.

\end{document}